\documentstyle[preprint,aps]{revtex}

\begin{document}
\draft
\author{Li-Bin Fu$^{1,}$\thanks{%
Email: fu$_-$libin@mail.iapcm.ac.cn }, Yishi Duan$^2$ and Hong
Zhang$^2$}
\address{$^1$ LCP, Institute of Applied Physics and Computational Mathematics, \\
P.O. Box 8009(26), Beijing 100088, P.R. China \\
$^2$ Physics Department, Lanzhou University, Lanzhou 730000, P.R. China}
\title{The Evolution of the Chern-Simons Vortices }

\maketitle

\begin{abstract}
\begin{center}
{\bf Abstract}
\end{center}

Based on the gauge potential decomposition theory and the $\phi $-mapping
theory, the topological inner structure of the Chern-Simons-Higgs vortex has
been showed in detail. The evolution of CSH vortices is studied from the
topological properties of the Higgs scalar field. The vortices are found
generating or annihilating at the limit points and encountering, splitting
or merging at the bifurcation points of the scalar field $\phi .$
\end{abstract}

\indent \indent \indent {\bf PACS number(s)}: 47.32.Cc, 11.15.-q, 02.40.PC


\newpage

\section{Introduction}

In resent years, a great deal of work on the Abelian Chern-Simons-Higgs
model in 2+1 dimension has been done by many physicists\cite{1,2,3,3a}. This
model has been widely used in many fields in physics, such as the fractional
spin in quantum field theory\cite{3,4}, and the quantum Hall effect in
condensed matter physics\cite{hall1,hall2}. Though it has been common to
include the topological properties of the Abelian CSH vortex, the
topological structure of this vortex has not been studied strictly. In this
paper, based on the decomposition theory of gauge potential and the $\phi $%
-mapping theory, the inner structure of the CSH vortex and its evolution
will be discussed in detail.

We know that the Abelian CSH Lagrangian density in 2+1 dimensions is often
expressed as
\begin{equation}
\label{csh1}L_{CSH}(\phi ,\,A)=\frac 14\alpha \in ^{\mu \nu \lambda }A_\mu
F_{\upsilon \lambda }+\frac 12D\phi (D\phi )^{*}+V(\phi ),
\end{equation}
where $\phi $ is the designated charged Higgs scalar field and $\frac 14%
\alpha \in ^{\mu \nu \lambda }A_\mu F_{\upsilon \lambda }$ is so-called
Chern-Simons term. As have been pointed out by many physicists\cite
{ykf,ashim}, the magnetic flux of the vortex is
\begin{equation}
\label{ch1}\Phi =\oint A_idx^i=\int \frac 12\in ^{ij}\partial _iA_jdx^2=%
\frac{2\pi \hbar c}en,
\end{equation}
where $n$ is a topological index, characterizing the vortex configuration. $%
\Phi $ is also referred to as the topological charge.

In our point of view\cite{DuanYangJiang,Duan2}, the $U(1)$ gauge potential
can be decomposed by the Higgs complex scalar field $\phi =\phi ^1+i\phi ^2$
as
\begin{equation}
A_\mu =\frac{\hbar c}e\in ^{ab}\partial _\mu n^an^b+\partial _\mu \lambda ,
\end{equation}
in which $\lambda $ is only a phase factor, and $n$ is a unit vector field
defined by%
$$
n^a=\frac{\phi ^a}{||\phi ||},\quad ||\phi ||=(\phi ^a\phi ^a)^{1/2}.
$$
It can be proved that there exists a topological current
\begin{equation}
\label{tpc}J^\mu =\frac 12\in ^{\mu \nu \lambda }\partial _\nu A_\lambda =%
\frac{\hbar c}{2e}\in ^{\mu \nu \lambda }\in _{ab}\partial _\mu n^a\partial
_\nu n^b.
\end{equation}
Obviously, the current (\ref{tpc}) is conserved. Following the $\phi $%
-mapping theory \cite{DuanYangJiang}, it can be rigorously proved that%
$$
J^\mu =\frac{2\pi \hbar c}e\delta ^2(\vec \phi )D^\mu (\frac \phi x),
$$
where the Jacobian $D^\mu (\frac \phi x)$ is defined as%
$$
\in ^{ab}D^\mu (\frac \phi x)=\in ^{\mu \nu \lambda }\partial _\nu
n^a\partial _\lambda n^b.
$$

According to the $\delta $-function theory \cite{dzf11} and the $\phi $%
-mapping theory, one can prove that
\begin{equation}
\label{curr}J^\mu =\frac{2\pi \hbar c}e\sum_{i=1}^l\beta _i\eta _i\delta ^2(%
\vec x-\vec z_i)\frac{dx^\mu }{dt}|_{z_i},
\end{equation}
where $\vec z_i$ are the $i$-th regular zero points of $\vec \phi ,$ i.e.
the solution of the following equation
\begin{equation}
\label{eq}\phi ^a(x)=0,\quad a=1,2,
\end{equation}
and the positive integer $\beta _i$ is the Hopf index\cite{yang28,y29,y30}
and $\eta _i=sign(D(\vec \phi /\vec x)_{z_i})=\pm 1$ is Brouwer degree\cite
{DuanGe,DuanZhang}. Then the density of topological charge can be wrote as
\begin{equation}
\label{densit}\rho =J^0=\frac{2\pi \hbar c}e\sum_{i=1}^l\beta _i\eta
_i\delta ^2(\vec x-\vec z_i).
\end{equation}
Comparing the formula to (\ref{ch1}) and considering Eq. (\ref{tpc}), the
total charge of the system given in (\ref{ch1}) can be rewrote as
\begin{equation}
\label{tal}Q=\int \rho (x)d^2x=\Phi _0\sum_{i=1}^l\beta _i\eta _i,
\end{equation}
where $\Phi _0=\frac{2\pi \hbar c}e$ is unit magnetic flux. And it is easy
to see that the topological index $n$ in (\ref{ch1}) has the following
expression
$$
n=\sum_{i=1}^l\beta _i\eta _i
$$
It is obvious to see that the vortex configuration given in (\ref{ch1}) is a
multivortex solution which possess the inner structure described by
expression (\ref{tal}). We can see vortex corresponds to $\eta _i=+1,$ while
anti-vortex corresponding to $\eta _i=-1.$

\section{The generation and annihilation of vortices}

As being discussed before, the zeros of the vector field $\vec \phi $ play
an important roles in describing the topological structure of the vortices.
Now we begin investigating the properties of the zero points. As we knew
before, if the Jacobian
\begin{equation}
D^0\left( \frac \phi x\right) \neq 0,
\end{equation}
we will have the isolated zeros of the vector field $\vec \phi .$ However,
when the condition fails, the above discussion will change in some way and
lead to the branch process. We denote one of the zero points as $(t^{*},\vec
z_i).$ If the Jacobian
\begin{equation}
\label{89}D^1(\frac \phi x)|_{(t^{*},\vec z_i)}\neq 0,
\end{equation}
we can use the Jacobian $D^1(\frac \phi x)$ instead of $D^0(\frac \phi x)$
for the purpose of using the implicit function theorem\cite{imp}. Then we
have a unique solution of the equations (\ref{eq}) in the neighborhood of
the limit point $(t^{*},\vec z_i)$
\begin{equation}
\label{92}t=t(x^1),\quad \quad x^2=x^2(x^1)
\end{equation}
with $t^{*}=t(z_i^1)$. We call the critical points $(t^{*},\vec z_i)$ the
limit points. In the present case, we know that
\begin{equation}
\label{wqd}\frac{dx^1}{dt}|_{(t^{*},\vec z_i)}=\frac{D^1(\frac \phi x)}{D(%
\frac \phi x)}|_{(t^{*},\vec z_i)}=\infty
\end{equation}
i.e.,
\begin{equation}
\frac{dt}{dx^1}|_{(t^{*},\vec z_i)}=0.
\end{equation}
Then, the Taylor expansion of $t=t(x^1)$ at the limit point $(t^{*},\vec z%
_i) $ is \cite{dzf5}
\begin{equation}
\label{93}t-t^{*}=\frac 12\frac{d^2t}{(dx^1)^2}|_{(t^{*},\vec z%
_i)}(x^1-x_i^1)^2
\end{equation}
which is a parabola in $x^1-t$ plane. From Eq.(\ref{93}) we can obtain two
solutions $x_1^1(t)$ and $x_2^1{}(t)$, which give two branch solutions
(worldlines of vortices). If $\frac{d^2t}{(dx^1)^2}|_{(t^{*},\vec z_i)}>0$,
we have the branch solutions for $t>t^{*}$ [see Fig. 1(a)]; otherwise, we
have the branch solutions for $t<t^{*}$ [see Fig. 1(b)]. These two cases are
related to the origin annihilate of the vortices.

One of the result of Eq. (\ref{wqd}), that the velocity of vortices are
infinite when they are annihilating, agrees with the fact obtained by Bray
\cite{Bray} who has a scaling argument associated with point defects final
annihilation which leads to large velocity tail. From Eq. (\ref{wqd}), we
also get a new result that the velocity of vortices is infinite when they
are generating, which is gained only from the topology of the scalar fields.

Since the topological current is identically conserved, the topological
charge of these two generated or annihilated vortices must be opposite at
the limit point, i.e.
\begin{equation}
\label{chargeI}\beta _{i_1}\eta _{i_1}=-\beta _{i_2}\eta _{i_2},
\end{equation}
which shows that $\beta _{i_1}=\beta _{i_2}$ and $\eta _{i_1}=-\eta _{i_2}$.
One can see that the fact the Brouwer degree $\eta $ is indefinite at the
limit points implies that it can change discontinuously at limit points
along the worldlines of vortices (from $\pm 1$ to $\mp 1$). It is easy to
see from Fig. 1: when $x^1>z_i^1$, $\eta _{i_1}=\pm 1$; when $x^1<z_i^1$, $%
\eta _{i_2}=\mp 1$.

For a limit point it is required that $D^1(\frac \phi x)|_{(t^{*},\vec z%
_i)}\neq 0.$ As to a bifurcation point\cite{dzf15}, it must satisfy a more
complex condition. This case will be discussed in the following section.

\section{The bifurcation of vortices velocity field}

In this section we have the restrictions of Eqs. (\ref{eq}) at the
bifurcation points $(t^{*},\vec z_i),$
\begin{equation}
\label{det-bif}D(\frac \phi x)|_{z_i}=0,\quad \quad \quad D^1(\frac \phi x%
)|_{z_i}=0,
\end{equation}
which leads to an important fact that the function relationship between $t$
and $x^1$ is not unique in the neighborhood of the bifurcation point $(t^{*},%
\vec z_i)$. It is easy to see that
\begin{equation}
\label{v-bif}V^1=\frac{dx^1}{dt}=\frac{D^1(\frac \phi x)}{D(\frac \phi x)}%
|_{z_i}
\end{equation}
which under (\ref{det-bif}) directly shows that the direction of the
integral curve of (\ref{v-bif}) is indefinite at $(t^{*},\vec z_i),$ i.e.,
the velocity field of vortices is indefinite at the point $(t^{*},\vec z_i)$%
. This is why the very point $(t^{*},\vec z_i)$ is called a bifurcation
point.

Assume that the bifurcation point $(t^{*},\vec z_i)$ has been found from
Eqs. (\ref{eq}) and (\ref{det-bif}). We know that , at the bifurcation point
$(t^{*},\vec z_i),$ the rank of the Jacobian matrix $[\partial \phi
/\partial x]$ is $1.$ In addition, according to the $\phi $-mapping theory,
the Taylor expansion of the solution of Eq. (\ref{eq}) in the neighborhood
of the bifurcation point $(t^{*},\vec z_i)$ can be expressed as \cite{dzf5}
\begin{equation}
A(x^1-x_i^1)^2+2B(x^1-x_i^1)(t-t^{*})+C(t-t^{*})^2=0
\end{equation}
which leads to
\begin{equation}
\label{bifb38}A(\frac{dx^1}{dt})^2+2B\frac{dx^1}{dt}+C=0
\end{equation}
and
\begin{equation}
\label{bifa39}C(\frac{dt}{dx^1})^2+2B\frac{dt}{dx^1}+A=0.
\end{equation}
where $A$, $B$ and $C$\ are three constants. The solutions of Eq. (\ref
{bifb38}) or Eq. (\ref{bifa39}) give different directions of the branch
curves (worldlines of vortices) at the bifurcation point. There are four
possible cases, which will show the physical meanings of the bifurcation
points.

Case 1 ($A\neq 0$): For $\Delta =4(B^2-AC)>0$\ from Eq. (\ref{bifb38}) we
get two different directions of the velocity field of vortices
\begin{equation}
\label{case1}\frac{dx^1}{dt}\mid _{1,2}=\frac{-B\pm \sqrt{B^2-AC}}A.
\end{equation}
\ which is shown in Fig. 2, where two worldlines of two vortices intersect
with different directions at the bifurcation point. This shows that two
vortices encounter and then depart at the bifurcation point.

Case 2 ($A\neq 0$): For $\Delta =4(B^2-AC)=0$\ from Eq. (\ref{bifb38}) we
get only one direction of the velocity field of vortices
\begin{equation}
\label{case2}\frac{dx^1}{dt}\mid _{1,2}=-\frac BA.
\end{equation}
\ which includes three important cases. (a) Two worldlines tangentially
contact, i.e. two vortices tangentially encounter at the bifurcation point
(see Fig. 3(a)). (b) Two worldlines merge into one worldline, i.e. two
vortices merge into one vortex at the bifurcation point (see Fig. 3(b)). (c)
One worldline resolves into two worldlines, i.e. one vortex splits into two
vortices at the bifurcation point. (see Fig. 3(c)).

Case 3 ($A=0,\,C\neq 0$): For $\Delta =4(B^2-AC)=0$\ from Eq. (\ref{bifa39})
we have
\begin{equation}
\label{case3}\frac{dt}{dx^1}\mid _{1,2}=\frac{-B\pm \sqrt{B^2-AC}}C=0,\quad -%
\frac{2B}C.
\end{equation}
\ There are two important cases: (a) One worldline resolves into three
worldlines, i.e. one vortex splits into three vortices at the bifurcation
point (see Fig. 4(a)). (b) Three worldlines merge into one worldline, i.e.
three vortices merge into one vortex at the bifurcation point (see Fig.
4(b)).

Case 4 ($A=C=0$): Eq. (\ref{bifb38}) and Eq. (\ref{bifa39}) gives
respectively
\begin{equation}
\label{case4}\frac{dx^1}{dt}=0,\quad \quad \frac{dt}{dx^1}=0.
\end{equation}
\ This case is obvious as Fig. 5 and is similar to case 3.

The above solutions reveal the evolution of the vortices. Besides the
encountering of the vortices, i.e. two vortices encounter and then depart at
the bifurcation point along different branch curves (see Fig. 2 and Fig.
3(a)), it also includes spliting and merging of vortices. When a
multicharged vortex moves through the bifurcation point, it may split into
several vortices along different branch curves (see Fig. 3(c), Fig. 4(a) and
Fig. 5(b)). On the contrary, several vortices can merge into one vortex at
the bifurcation point (see Fig. 3(b), Fig. 4(b)).

From Eqs. (\ref{case3}) and (\ref{case4}), we find that there are two
branches includes the case that the velocity of vortices is infinite at the
bifurcation point, i.e. when one vortex splits into three vortices or three
merge into one, there must exists a pair of vortices whose velocity is
infinite at the bifurcation point. It seems that these cases are associated
with the process of generating or annihilating. This is pointed to be
verified in future.

The identical conversation of the topological charge shows the sum of the
topological charge of these final vortices must be equal to that of the
original vortices at the bifurcation point, i.e.
\begin{equation}
\label{chargeII}\sum_i\beta _{l_i}\eta _{l_i}=\sum_f\beta _{l_f}\eta _{l_f}
\end{equation}
for fixed $l$. Furthermore, from above studies, we see that the generation,
annihilation and bifurcation of vortices are not gradual change, but sudden
change at the critical points.

\section{Conclusion}

Firstly, we obtain the inner topological structure of Chern-Simons vortex.
The multi-charged vortex has been found at the every zero point of the Higgs
scalar field $\phi $ under the condition that the Jacobian determinate $%
D(\phi /x)\neq 0$. One also shows that the charge of the vortex is
determined by Hopf indices and Brouwer degrees. Secondly, we conclude that
there exist crucial cases of branch processes in the evolution of the
vortices when $D(\frac \phi x)=0$, i.e., $\eta _i$ is indefinite. This means
that the vortices generate or annihilate at the limit points and encounter,
split or merge at the bifurcation points of the Higgs scalar fields, which
shows that the vortices system is unstable at these branch points. Here we
must point out that there exist two restrictions of the evolution of
vortices. One restriction is the conservation of the topological charge of
the vortices during the branch process (see Eqs. (\ref{chargeI}) and (\ref
{chargeII})), the other restriction is the number of different directions of
the worldlines of vortices is at most four at the bifurcation points (see
Eqs. (\ref{bifb38}), (\ref{bifa39}) ). Perhaps, the former is known before,
but the later is pointed out for the first time.

\section*{Figure Captions}

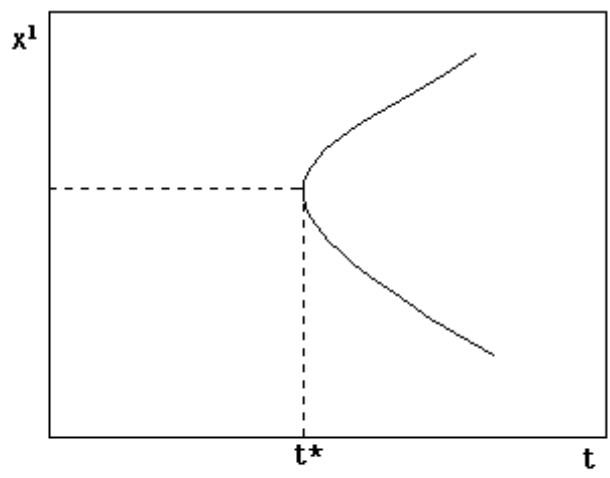
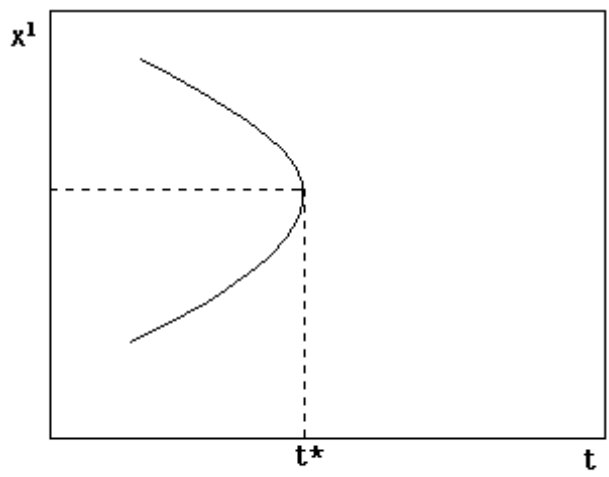
Fig. 1. Projecting the worldlines of vortices onto ($x^1-t$)-plane. (a) The
branch solutions for Eq. (\ref{93}) when $\frac{d^2t}{(dx^1)^2}|_{(t^{*},%
\vec z_i)}>0$, i.e. a pair of vortices with opposite charges generate at the
limit point, i.e. the origin of vortices. (b) The branch solutions for Eq. (%
\ref{93}) when $\frac{d^2t}{(dx^1)^2}|_{(t^{*},\vec z_i)}<0$, i.e. a pair of
vortices with opposite charges annihilate at the limit point.

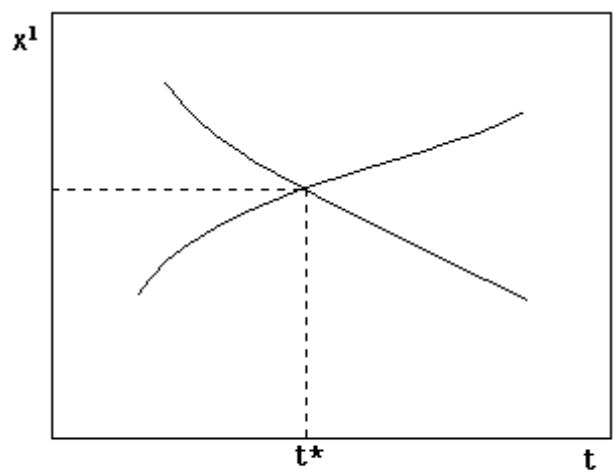
Fig. 2. Projecting the worldlines of vortices onto ($x^1-t$)-plane. Two
worldlines intersect with different directions at the bifurcation point,
i.e. two vortices encounter at the bifurcation point.

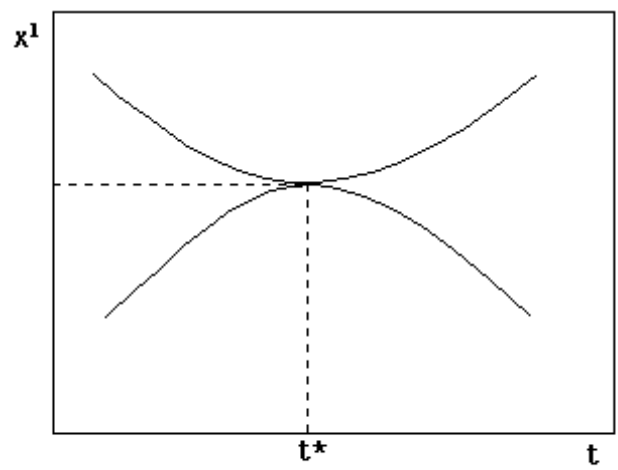
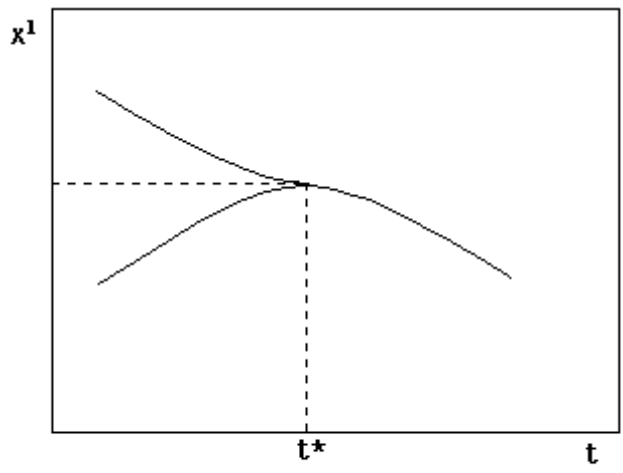
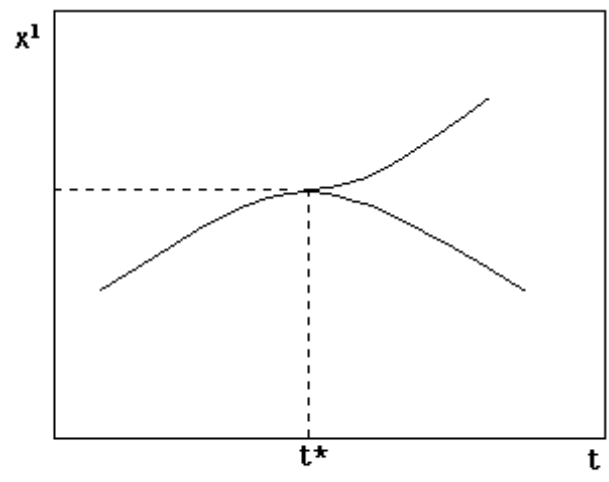
Fig. 3. (a) Two worldlines tangentially contact, i.e. two vortices
tangentially encounter at the bifurcation point. (b) Two worldlines merge
into one worldline, i.e. two vortices merge into one vortices at the
bifurcation point. (c) One worldline resolves into two worldlines, i.e. one
vortex splits into two vortices at the bifurcation point.

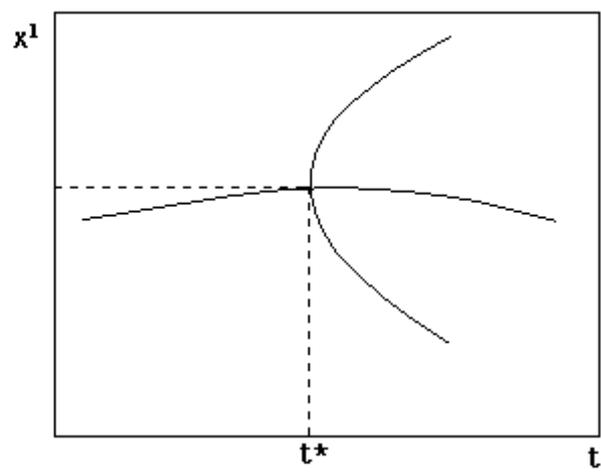
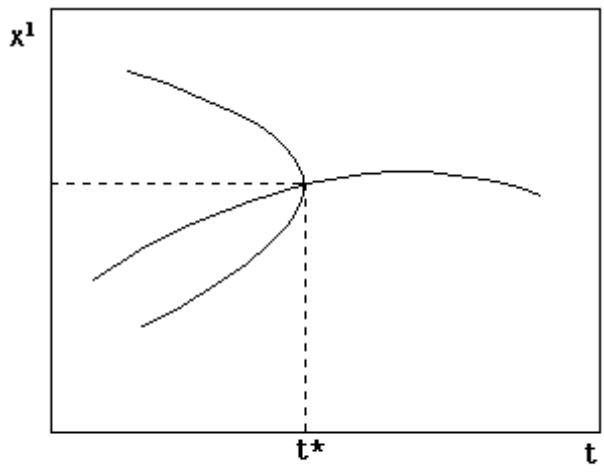
Fig. 4. Two important cases of Eq. (\ref{case3}). (a) One worldline resolves
into three worldlines, i.e. one vortex splits into three vortices at the
bifurcation point. (b) Three worldlines merge into one worldline, i.e. three
vortices merge into one vortex at the bifurcation point.

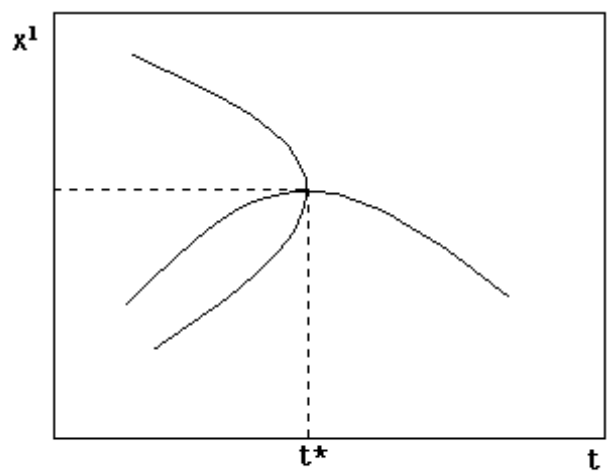
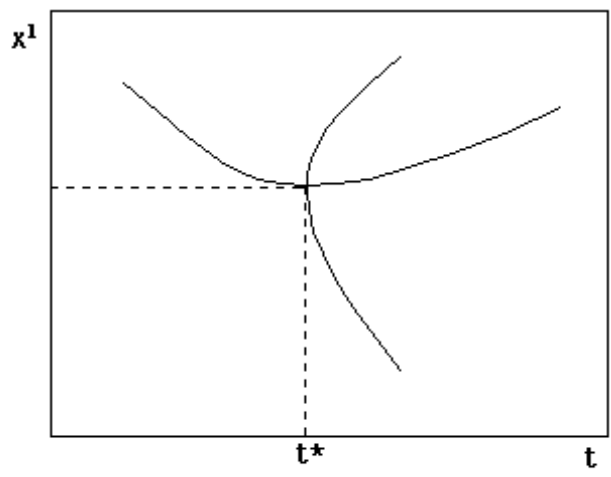
Fig. 5. Two worldlines intersect normally at the bifurcation point. This
case is similar to Fig. 4. (a) Three vortices merge into one vortex at the
bifurcation point. (b) One vortex splits into three vortices at the
bifurcation point.

\end{document}